\documentstyle[prl,multicol,aps]{revtex}

\begin{document}
\input{epsf.tex}
\epsfverbosetrue

\title{Transition radiation by matter-wave solitons in optical lattices}
\author{A.V. Yulin, D.V. Skryabin, and P.St.J. Russell}
\address{Department of Physics, University of Bath, Bath BA2 7AY, United
Kingdom}

\maketitle
\begin{abstract}
We demonstrate that matter-wave  solitary pulses formed from
Bose condensed atoms moving inside optical lattices continuously radiate dispersive
matter waves with prescribed momentum.
Our analytical results for the radiation parameters and the soliton decay rate
are found to be in excellent agreement
with numerical modelling performed for experimentally relevant parameters.
\end{abstract}
\begin{multicols}{2}
\narrowtext

Recent observations  of matter-wave solitons \cite{dark,bright}
have clearly been amongst the most breaking achievements in the
burgeoning  field of  Bose-Einstein condensation of dilute atomic
gases. Balance between the spatial dispersion of matter waves  and
repulsive or attractive interatomic interactions ensures existence
of dark \cite{dark} or bright \cite{bright}
solitons, respectively. Dispersion of the atomic condensates can, however, be
reversed by embedding the condensate into a periodic potential
created by standing light waves, i.e. {\em optical lattice}
\cite{lattice,motion,erman}.
The idea  of changing the dispersion  sign and of the
possible observation of the bright matter-wave solitons
in the condensates with repulsive interatomic interaction
has been around for a while, see \cite{wright} and references therein, and
more theoretical results
have been produced recently, see, e.g., \cite{markus,kivshar,scott},
in the view of the
rapid maturing of the experimental techniques \cite{lattice,motion,erman}.
The concept of the dispersion control by periodic potentials
is also well known in solid-state physics \cite{bloch_book}
and a very active topic of research in nonlinear optics, see, e.g., \cite{ben}.

To understand the initial motivation which leads to the results described below,
it is instructive to recall the effect of the {\em transition radiation} known
from classical electrodynamics \cite{landau}. Transition radiation is a continuous emission of
electromagnetic waves by a charged particle
moving with a {\em constant} velocity in spatially inhomogeneous medium.
This radiation is emitted because the field created by
the particle has different characteristics in different parts of the medium. When the
particle moves, the field reorganizes itself continuously and
shakes off some of its parts in form
of the radiation.  We expect that similar phenomenon should take place with
matter-wave solitons in optical lattices.

In order to change the dispersion  sign of matter waves forming
the initially resting one-dimensional packet embedded into the
optical lattice, one needs to  position it in  momentum
space somewhere between the inflexion point of the
energy-momentum, i.e., dispersion, characteristic  and the edge of
the Brillouin zone, see Fig. 1(a) \cite{erman1}.
Only  exactly at the edge of the zone
does the group velocity go to zero. Solitons
with a spread of  quasi-momenta centered at the edge
will therefore be the only resting bright solitons in the condensates with
repulsive inter-atomic interaction. Our primary interest below is,
however, moving solitons. Once  a solitonic wave packet moves
through the periodic potential one can expect that its structure
will not be able to  instantly readjust itself to perfectly fit
the conditions that the local density maxima are positioned at the
center of the local potential minima.
Therefore the moving soliton has to continuously overcome the energy barrier,
which leads to the continuous emission of matter waves.
This energy barrier is analogous to the Peierls-Nabarro potential
known for the solitons in discrete systems \cite{nabarro}.
Using analogy with the
electromagnetic case we term this effect as transition radiation of matter waves.

We start our analysis from the Gross-Pitaevskii (GP) equation describing evolution of
the macroscopic wave function  of the zero-temperature Bose-Einstein condensate  (BEC)
interacting with off-resonant standing light wave. We assume that the condensate
is tightly confined by the external harmonic potential along the $Y$ and $Z$ directions
having the trap frequencies $\omega_{Y,Z}=2\pi\times 400 s^{-1}$
and that  any deviations of the  potential along $X$-axis from the $\sin^2k_lX$
produced by the intensity of
the standing laser field with wavenumber $k_l$ can be disregarded.
We take for our estimates that $k_l=2\pi/800 nm^{-1}$ and consider BEC  made of $^{87}Rb$ atoms
with a two-body scattering length $a\simeq 5.4\times 10^{-9}m$.
The characteristic transverse width of this BEC  is then given by
$w=\sqrt{\hbar/(m\omega_{Y})}\simeq 0.5\mu m$.
Assuming that the condensate profile along the $Y,Z$
directions is given by the lowest mode of the harmonic potential,
we can derive the one-dimensional GP equation, which  describes dynamics of
the $X$ dependent part of the full wave function.
The dimensionless normalized form of this equation is \cite{kivshar}:
\begin{equation}
i\partial_{t}\psi=-\partial_{x}^2\psi-
\beta\psi\cos 2x+|\psi|^2\psi. \label{1_1}
\end{equation}
Here the dimensionless time $t$ and spatial coordinate $x$ are
measured respectively in the units of $T_0=2m/(\hbar k_l^2)\simeq
5\times 10^{-5}s$ and $1/k_l\simeq 0.13\mu m$. Meaning of the
dimensionless parameter $\beta$ is easily inferred from the
expression for the  lattice potential in the physical units, which
is taken as  $4\beta E_r\sin^2k_lX$, where
$E_r=\hbar^2k_l^2/(2m)\simeq\hbar\times20 kHz$ is the recoil
energy. To estimate number of atoms $N$ in the condensate we
introduce the effective area  in the $(Y,Z)$-plane, $A_{eff}=2\pi
w^2\simeq 1.5\mu m^2$, and the atom density $n\simeq
10^{14}cm^{-3}$. Then $N=\int  |\psi|^2 dx\times
A^2_{eff}k_ln^{2/3}/(8\pi |a|) \simeq 4\times 10^3\times\int
|\psi|^2 dx$.

We proceed by expanding $\psi$
over the Bloch functions, $b(x,k)$
\cite{bloch_book}: $\psi(x,t)=\int dk \tilde \psi(k,t)b(x,k)$.
Here $b(x,k)$ are eigenfunctions of the operator
$\hat L=\partial_x^2+\beta\cos 2x$, such that $\hat Lb=-\epsilon(k)b$,
where $k$ is the quasi-momentum introduced as
 $b(x,k)=g(x,k)e^{ixk}$ \cite{bloch_book},   $\epsilon(k)$ is
the energy of the non-interacting, i.e. linear, matter-waves,
and $g(x,k)$ is the function with the spatial period $\pi$.
The first allowed, first forbidden (gap) and a small part of the 2nd allowed
energy bands are  shown in the energy-momentum plot in Fig. 1(a).

Choosing  $k=k_s\in [-1,1]$,   we  expand
$g(x,k)$ in a Taylor series around $k=k_s$ and demonstrate  that
\begin{equation}
\psi(x,t)=e^{ik_sx}\hat G_sA(x,t).\label{d_2}
\end{equation}
Here, $A(x,t)=\int dk \tilde\psi(k,t)e^{i(k-k_s)x}$ and $\hat G_s$ is the
linear differential operator:
$\hat G_s=\sum_{n=0}^{\infty}(1/n!)\partial^n_kg(x,k_s)(-i\partial_x)^n$.
Similarly $\hat L\psi$ can be represented as
\begin{equation}
\hat L\psi=-e^{ik_sx}\hat G_s\hat{\cal E}_s[-i\partial_x] A(x,t),\label{d_3}
\end{equation}
where $\hat{\cal E}_s[-i\partial_x]=
\sum_{m=0}^{\infty}(1/m!)\partial_{k}^m\epsilon(k_s)(-i\partial_x)^m$
is  the energy operator. Subscript '$s$' refers to
quantities and functions calculated for $k=k_s$.

After substitution of Eqs. (\ref{d_2}), (\ref{d_3}) into Eq. (\ref{1_1}), we then
replace $\hat G_s$ with its 1st order approximation $g(x,k_s)\equiv g_s$ and
$\hat{\cal E}_s$ with $\hat{\cal E}_{2s}\equiv\epsilon_s-i\epsilon^{\prime}_s\partial_x
-{1/2}\epsilon^{\prime\prime}_s\partial_x^2$. Here $\epsilon^{\prime}_s$ and
$\epsilon^{\prime\prime}_s$ are, respectively, group velocity and group velocity dispersion
of matter-waves. Assuming smallness of the nonlinearity and
averaging the resulting equation over $g_s$ we  derive the
renowned nonlinear Schr\"odinger (NLS) equation
$i\partial_tA=\hat{\cal E}_{2s}A+\alpha_s|A|^2A$,
where $\alpha_s=\int dx|g_s|^4$.
 Below we are interested
in the bright solitons  with $\epsilon_s$ belonging to the first allowed energy band. These are given by
$A_s=R_s(\xi)\exp\{-i\kappa t-i\epsilon_s t\}$, where
\begin{equation}
 R_s(\xi)=\sqrt{2\kappa\over\alpha_s}
sech\left\{\xi\sqrt{2\kappa\over -\epsilon_s^{\prime\prime}}
\right\},~\xi=x-\epsilon_s^\prime t,\label{soliton}
\end{equation}
$sgn\kappa=sgn\alpha_s=-sgn\epsilon^{\prime\prime}_s$ and
$\kappa\ne 0$ is the nonlinearity induced energy shift. Details of the derivation of NLS equation
 from Eq. (\ref{1_1}) has been previously published
using the method of  multiple scales \cite{mario}.
Approximations made above and in \cite{mario} imply that $\psi$ has sufficiently narrow
spread of quasi-momenta around $k=k_s$. However, the
approach introduced here is readily adaptable to  give an  access to
the small amplitude corrections having quasi-momenta detuned
 far  from $k_s$, see Eqs. (8), (9) below. Note, here that mobile
 envelope solitons (\ref{soliton})
 are very different in their properties from practically
 immobile  solitons occupying primarily one or few lattice sites and considered, e.g.,
 in \cite{kivshar}.

To introduce the effect studied and explained in this work, we first present
the results of  numerical modelling of Eq. (\ref{1_1}) with initial conditions
in the form $g_s(x)R_s(x)e^{ik_sx}$ for  values of
 $k_s\in (k_0, 1]$, $\epsilon^{\prime\prime}(k_0)=0$, ensuring that the effective mass,
$1/\epsilon^{\prime\prime}_s$, is negative, see Fig. 1(a). Taking
$k_s=1$, i.e. fixing soliton parameters at the point corresponding
to the zero of the group velocity, we observe formation of
the ideal solitary pulse \cite{markus}. The Fourier spectrum of this
solution contains series of the equidistant peaks, the location of
which is determined by the spectrum of the corresponding Bloch
function $b(x,k_s=1)$. Note, that the wave-numbers $q$,
parameterizing the Fourier transform of $\psi$, $\psi=\int
dx\tilde\Psi(q)e^{iqx}$, are linked with quasi-momenta $k$ as
$q=k\pm 2n$. Here and below $n=0, 1, 2,\dots$ numbers the Brillouin zones.
For  values of $k_s\ne 1$  we have observed the quasi-solitonic
pulses, which, while travelling, leave behind the trail of small
amplitude radiation, Fig. 2(a). The radiation effect becomes
noticeably stronger for $k_s\to k_0$. Spectra of the radiating
solitons have a distinct peak, see Fig. 2(b), which is absent in
the spectra of the ideal resting solitons. The overall results of the
extensive series of numerical experiments unambiguously indicate
that solitary pulses moving through the optical lattice
continuously emit radiation with certain spectrally localized
quasi-momenta. By analogy with electromagnetic case we term this
radiation as {\em transition radiation of matter waves}.

The  initial conditions used above in the form of the $sech$ envelope superimposed
on the Bloch function $g_s(x)e^{ik_sx}$ are difficult, though probably not impossible
to prepare in the real experiments. However, current experimental techniques
allow straightforward preparation of the gaussian matter-wave packets
with spectrum centered around the zero quasi-momentum and setting
lattices in motion. The  lattice moving with velocity $2k_s$ will then effectively
shift the central momentum of the wave packet to $k_s$.
$2k_s$ equals to the group velocity of the free,
i.e. without the lattice, matter waves with quasi-momentum $k_s$.
In  turn,
the backward scattering of matter waves from the moving lattice is expected
to create the second strong and other peaks in the spectrum of the wave-packet.
The results of modelling of Eq. (1) with moving potential $\cos 2(x-2k_st)$
and initial conditions in the form of the gaussian packet are shown in Figs. 2(c,d).
Taking into account shift of the axes, one can see that
these results are in  remarkable agreement with those obtained
using the solitonic initial conditions, see Figs. 2(a,b).
Good resolution of the radiation peaks in Figs. 2(b,d) suggests even that they
can possibly be recorded experimentally.

To understand and give analytical interpretation of the observed
radiation we develop a perturbative approach, allowing us to
predict  quasi-momenta and   the amplitude of the emitted wave and,
thereby allowing an estimate of the decay rate of a solitary pulse.
To proceed we form the ansatz
\begin{equation}
\psi=\psi_s(x,\xi,t)+\varphi(t,x),\label{rad_1}
\end{equation}
where the first term   approximates the solitonic
part of the wave function, $\psi_s=g_s(x)A_s(\xi,t)e^{ik_sx}$,
and the second one is an  arbitrary perturbation.
Substituting   (\ref{rad_1}) into  (\ref{1_1}) and assuming that $|\varphi|$ is small
we find that evolution of  $\varphi$ is governed by
\begin{equation}
i\partial_{t}\varphi+\hat L\varphi - 2|\psi_s|^2\varphi - \psi_s^2
\varphi^{*}= S(x,t), \label{y_1}
\end{equation}
where $S(x,t)=-i\partial_{t}\psi_s-\hat L\psi_s + |\psi_s|^2\psi_s$
is the source term, which
is different from zero because $\psi_s$ is not an exact solution of Eq. (\ref{1_1}).

The  structure of the radiation
tail observed in the numerical modelling  corresponds to
the spatially extended lattice eigenmode.
The natural mechanism for excitation of a selected eigenmode from the continuum is  the
energy and wave-number resonance with one  Fourier component of the source term $S$.
The resonance  condition ensures that
there exists a lattice mode which is always in-phase and therefore interferes
constructively with one Fourier component
of the source term. The latter can be represented in the form
 $S=e^{ik_sx-i\epsilon_s t-i\kappa t}\sum_{j}h_j(x)f_j(\xi)$,
 where the sum is taken over all the terms appearing in
 the right-hand side after substitution of the explicit expression for the soliton and
 $h_j(x)$ are some functions with a period of $\pi$.
 Replacing $f_j(\xi)$ through their Fourier integrals  $f_j(\xi)=\int dQ \tilde f_j(Q)e^{iQ\xi}$
 and recalling that $\xi=x-\epsilon^{\prime}_st$,
 one can easily find that the lattice modes satisfying the
 resonance condition are those, which have quasi-momenta $k=k_r$
obeying the condition $\epsilon(k_r) = \kappa+\epsilon_s + \epsilon_s^{\prime} Q_r$.
$Q_r=Q(k_r)$  is the detuning
of the wave-number of the resonant wave
$q_r=k_r\pm 2n$ from  the soliton quasi-momentum $k_s$, i.e. $Q_r=k_r\pm 2n-k_s$.
Using that $\epsilon(k_r)=\epsilon(q_r)$ the resonance
condition can be rewritten in the form
\begin{equation}
\epsilon(q_r) = \kappa+\epsilon_s + (q_r-k_s)\epsilon_s^{\prime}.
\label{graph}
\end{equation}
The geometrical meaning of Eq. (\ref{graph}) is  clear. The
right-hand side of (\ref{graph}) equals to
the energy of the  Fourier component of the soliton (\ref{soliton}), while its left hand
side is simply the energy of the linear dispersive wave.
Eq. (\ref{graph}) can be  solved for $q_r$ by plotting the
tangent line to the periodically extended dispersion characteristic
at the point $\epsilon=\epsilon_s$. Then one should make the parallel upshift of this tangent
by $\kappa$ and find  points of the intersection with the
dispersion characteristic itself. Topologically it is self-evident that only
resting solitons do not produce any real roots of Eq. (\ref{graph}) and do not radiate
into the lattice modes. For $\epsilon_s^{\prime}\ne 0$
Eq. (\ref{graph}) has infinitely many real roots corresponding to different values of $n$.
Root $q_r$ for $n=0$, i.e. when $q_r=k_r$, as a function of $k_s$
is shown in Fig. 1(b). Dots on this graph show
$(k_r,k_s)$  pairs measured from the direct numerical modelling of Eq. (1).
Modelling of Eq. (\ref{1_1}) has not revealed any  resonances in the Brillouin zones with $n\ne 0$.
It indicates that coupling into the higher-order resonances is
negligible, primarily because spectral strength of the source term is very weak
for the values of $Q_r$ with $n\ne 0$.
Note, that the secondary less intense spectral peaks, seen in Figs. 2(b,d),
 are described by the peaks in the Fourier spectra of $b(x,k_{r,s})$,
 and are given by $q_{r,s}=k_{r,s}\pm 2n$. At the same time,
 multiple roots of Eq. (\ref{graph}) are not linked by any simple algebraic expression.

In order to  calculate the amplitude of the radiated wave
we form the anzats
\begin{equation}
\psi\simeq \psi_s(x,\xi,t)+g_r(x)e^{ik_rx-i\epsilon_r t}W(t,\xi),\label{rad_2}
\end{equation}
where subscript '$r$' refers to the quantities calculated at $k=k_r$.
Substituting (\ref{rad_2}) into Eq. (\ref{1_1}) we take into account
that $\hat Lg_r(x)e^{ik_rx}W(t,\xi)\simeq -g_r(x)e^{ik_rx}\hat{\cal E}_{r}W(\xi,t)$.
The latter expression can be easily inferred by comparison with Eq. (\ref{d_3}).
Assuming that $\delta k_s\ll |k_s-k_r|\ll 2$,
where $\delta k_s$ is the width of the soliton in the
quasi-momentum space and $2$ is the width of the Brillouin zone,
and disregarding terms nonlinear in W, one  gets the averaged equation for the  amplitude of
the radiation field:
$\vec  s=i\partial_t\vec W+ \hat{\cal L}_r \vec W$,
 where $ \vec s =(  P ,-  P^* )^T$, $\vec W=(W,W^*)^T$,
$P =e^{i\xi (k_s-k_r)}[\alpha_4\{\hat {\cal E}_s
-\hat {\cal E}_{2s}\}R_s+\alpha_3|R_s|^2R_s]$,
$\hat D_r[-i\partial_{\xi}]=-\epsilon_r+
i\epsilon_s^{\prime}\partial_{\xi}+\hat{\cal E}_r[-i\partial_{\xi}]$,
\begin{eqnarray}
  && \hat{\cal L}_r =\left[\begin{array}{cc}
-\hat D_r+\alpha_2|R_s|^2&\alpha_1R_s^2e^{2i\xi(k_s-k_r)}\\
-\alpha_1^*R_s^2e^{2i\xi(k_r-k_s)} &\hat D^*_r-\alpha_2^*|R_s|^2
\end{array}\right],\label{operator}
\end{eqnarray}
 $\alpha_1=\int dxg_s^2{g_r^*}^2$,
$\alpha_2=2\int dxg_s^2|g_r|^2$, $\alpha_3=\alpha_s\int dxg_sg_r^*- \int dx|g_s|^2g_sg_r^*$,
and $\alpha_4=\int dxg_sg_r^*$.

The approximate solution we obtain  for $W$ is
\begin{equation}
W(t,\xi)\simeq -iC\left\{
\Theta(\zeta\xi)-\Theta(\zeta\xi+\zeta t(\epsilon_s^{\prime}-\epsilon_r^{\prime}))
\right\}. \label{ampl}
\end{equation}
Here $\zeta=sgn(\epsilon_r^{\prime}-\epsilon_s^{\prime})$,
$\Theta$ is the Heaviside function and $C$ is the amplitude, which
can not be generally expressed in a closed analytical form and, therefore,
was calculated numerically. $C$ characterizes the spectral intensity of the
source term for $k=k_r$. Heaviside functions in Eq. (\ref{ampl}) describe the tail of
the radiation field having the length $t|\epsilon_r^{\prime}-\epsilon_s^{\prime}|$. The tail
starts at the soliton ($\xi=0$) and
extends beyond ($\epsilon_r^{\prime}<\epsilon_s^{\prime}$) or in
front ($\epsilon_r^{\prime}>\epsilon_s^{\prime}$) of it.

Substituting the ansatz  (\ref{rad_2}) into the conservation law $\partial_t\int dx|\psi|^2=0$,
we can estimate the rate, $\gamma$, of the transfer
of  particles from the soliton to the  radiation:
$\gamma=C^2|\epsilon_s^{\prime}-\epsilon_r^{\prime}|$.
Plot of $\log_{10}\gamma$ vs $k_s$ for two values of $\kappa$ is shown in Fig. 3(a).
Naturally, the rate of transfer increases, when detuning $|k_r-k_s|$ decreases for
$k_s\to k_0$, see Fig. 1(b). This is because for $k_s\to k_0$ the lattice mode is  resonant
with the most  intense central part of the soliton spectrum.
Contrary, $|k_r-k_s|$ increases for $k_s\to 1$, and
the radiation amplitude decays almost exponentially.
For example, semi-analytical Eq. (\ref{ampl}) gives that
the initial condition with $k_s=0.87$
used to generate Fig. 2 has the initial decay rate $3\times 10^4$
particles per second. Providing that this rate is constant in time,
the  soliton life-time would be  $\simeq 0.033$s.
However,  radiation carries away both density and momentum from the solitonic
part of the field. Therefore solitonic parameters $\kappa$ and $k_s$ are in fact
functions of time. In particular, we have found
that the radiation results in  convergence
of the soliton quasi-momentum to some limit value, which is always closer to $1$
than the initial $k_s$. Thus, radiation emission slows  the soliton.
For example,  taking initial conditions with $\kappa=0.01$ and $k_s=0.87$
we have observed that over  $0.05$s  ($1000$ dimensionless time units)
$k_s$ shifts  to $\simeq 0.887$, see Fig. 3(b), and $\kappa$  to $\simeq 0.006$.
The relatively small increase of $k_s$ is accompanied by the decrease of the soliton decay
rate, which drops, accordingly with plots shown in Fig. 3(a),
from  $3\times 10^4$ to $\simeq 10^2$ particles per second.
Let us stress again that the soliton life-time tends to infinity when $k_s\to 1$.
Correspondingly, the estimated soliton life-time increases dramatically to $\simeq 10$s.
Fig. 3(b) also shows the numerically computed slowdown in the growth  of the number of particles
in the radiation   component of the condensate. Soliton decay rate inferred from these data
and corresponding theoretical points calculated for instant values of the soliton
parameters $\kappa$ and $k_s$ are in good agreement, see inset in Fig. 3(b),
which  confirms validity of our
theoretical method.

In summary, we have reported  transition radiation by matter-wave
solitons moving through the optical lattice. This effect extends
family of the already known and related quantum radiative  effects
such as, e.g., sound emission by  precessing quantized vortices
\cite{pismen} and   by  dark matter-wave solitons oscillating in a
harmonic trap \cite{adams}. Note, also that our main conclusions
and techniques can be used to predict and analyze radiation by
spatial optical solitons moving in  nonlinear photonic crystals
\cite{ben}. This work was partially supported by the INTAS project
211-855.



\begin{figure}
\setlength{\epsfxsize}{10.0cm}
\centerline{\epsfbox{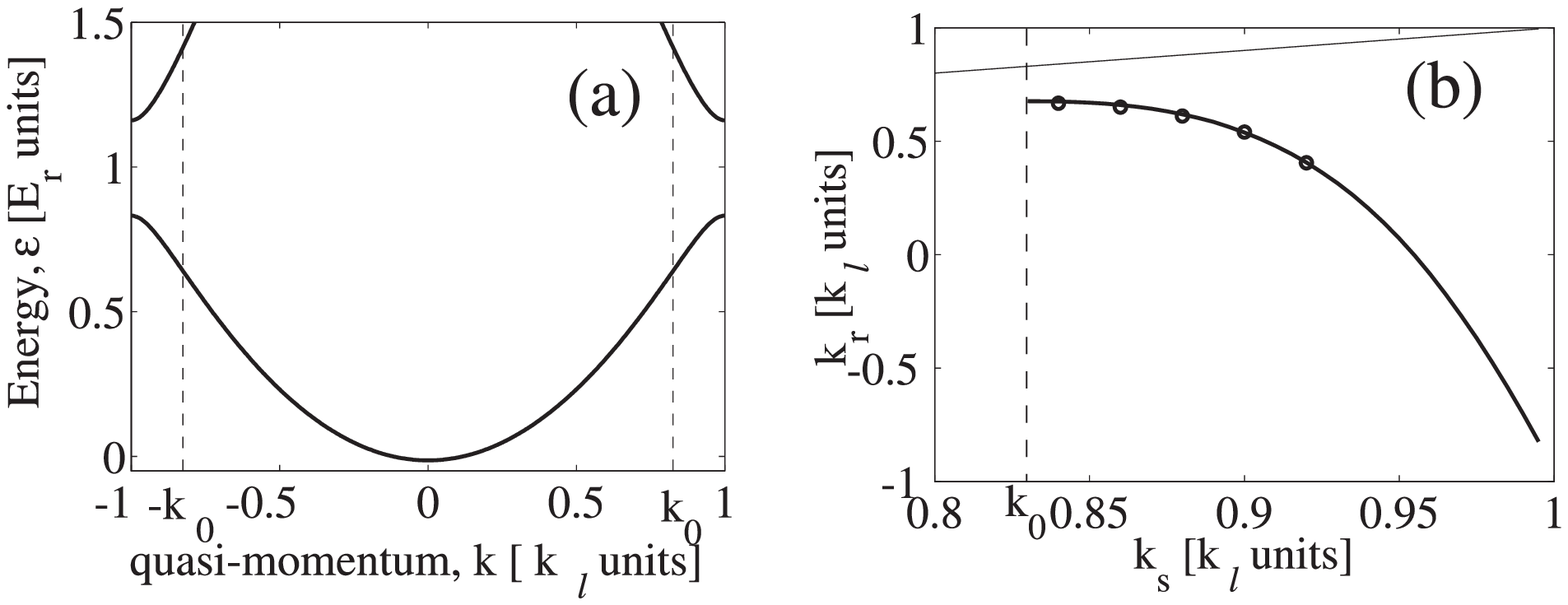}}
\caption{(a) Energy-momentum diagram for the linear matter-waves in optical lattice:
$n=0$ Brillouin zone is shown.
(b) Dependence of the  radiation quasi-momentum, $k_r$,
from the soliton quasi-momentum, $k_s$. Full diagonal line in (b) corresponds to $k_r=k_s$.
Dots in (b) mark  $(k_s,k_r)$ pairs measured from the modelling of Eq. (\ref{1_1}).
Dashed vertical lines in (a) and (b) mark the $\pm k_0$ points with $\epsilon^{\prime\prime}=0$.
$\beta=0.33$}
\label{fig1}\end{figure}

\begin{figure} \setlength{\epsfxsize}{10.0cm}
\centerline{\epsfbox{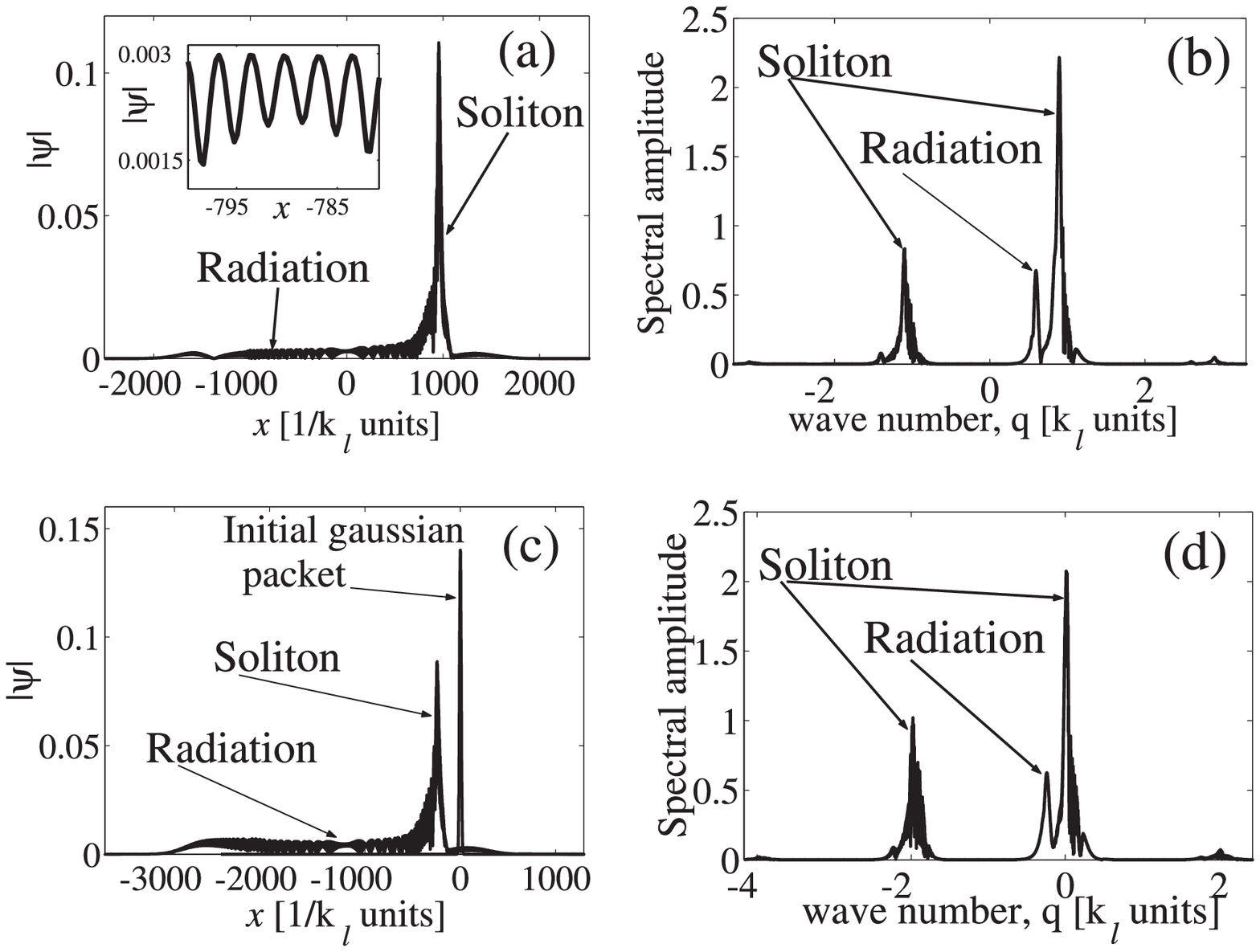}}
\caption{Results of the direct numerical modelling of Eq.
(\ref{1_1}). (a,c) Square roots of the atomic density as functions
of $x$. The inset in (a) shows fine details of the spatial profile
of the radiation. (b,d) Corresponding Fourier spectra. (a,b)  are
obtained for the initial  solitonic wave packet with $k_s=0.87$
and $\kappa=0.01$. (c,d) are obtained for the initial gaussian
wave packets with the zero central momentum and lattice moving
with velocity $2k_s=1.74$. Integration time $t=700$ corresponds to
$0.035$s. The physical number of particles in the  soliton shown
in Fig. 2(a) is $\simeq  10^3$,  its velocity is $3\times
10^{-3}m/s$ and  width at the half-height is $\simeq 4\mu m$.}
\label{fig2}\end{figure}

\begin{figure} \setlength{\epsfxsize}{10.0cm}
\centerline{\epsfbox{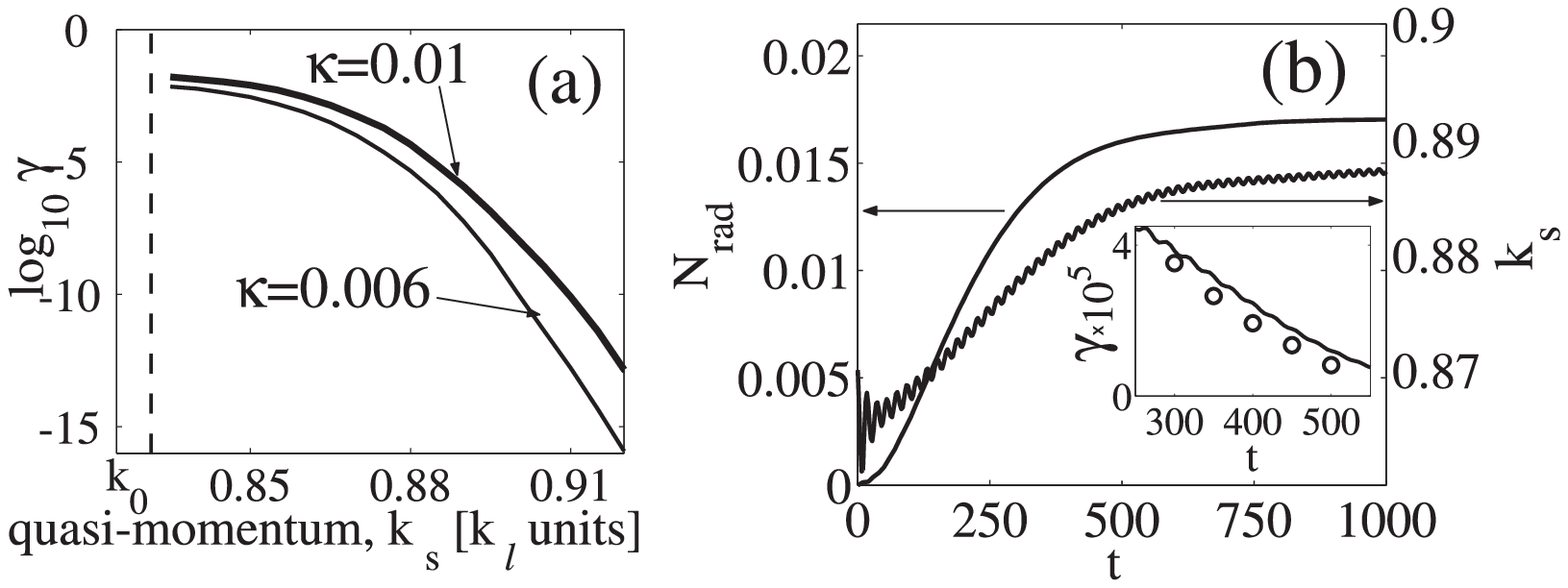}}
\caption{(a) Logarithm of the particle transfer rate from the solitonic to the radiation
part of the wave function as a function of the soliton quasi-momentum for $\kappa=0.01$ and $0.006$.
Dashed vertical line marks the  $\epsilon^{\prime\prime}=0$ point.
(b) Temporal evolution of the normalized number of particles $N_{rad}$
in the radiation component of the field (left axis) and corresponding dynamics of
the soliton quasi-momentum (right axis). $N_{rad}$ is calculated as  integral of $|\psi|^2$
over the tail behind the soliton. Estimate for physical number of particles in the radiation
tail is given by $3\times 10^3N_{rad}$. Inset shows the soliton decay rate $\gamma$ as function of $t$:
solid line -- numerical modelling and dots -- theoretical results.} \label{fig3}\end{figure}

\end{multicols}

\begin{references}

\bibitem{dark}
S. Burger {\em et. al.}, Phys. Rev. Lett.  {\bf  83}, 5198 (1999);
J. Denschlag {\em et. al.}, Science {\bf 287}, 97 (2000).


\bibitem{bright}
K.E. Strecker {\em et. al.}, Nature (London) {\bf 417}, 150 (2002);
L. Khaykovich {\em et. al.}, Science {\bf 296}, 1290 (2002).

\bibitem{lattice}
B.P. Anderson and M.A. Kasevich, Science {\bf 282}, 1686 (1998);
F.S. Cataliotti {\em et. al.}, Science {\bf 293}, 843 (2001);
M. Greiner {\em et. al.}, Nature (London) {\bf 419}, 51 (2002);
C. Fort {\em et. al.}, Phys. Rev. Lett.  {\bf  90}, 140405 (2003).

\bibitem{motion}
S. Burger {\em et. al.}, Phys. Rev. Lett.  {\bf  86}, 4447 (2001);
O. Morsch {\em et. al.}, Phys. Rev. Lett.  {\bf  87}, 140402 (2001);
J.H. Denschlag {\em et. al.},  J. Phys. B {\bf 35}, 3095 (2002).

\bibitem{erman}
B. Eiermann {\em et al.}, Phys. Rev. Lett. {\bf 91}, 060402 (2003).

\bibitem{nabarro}
Y.S. Kivshar and D.K. Campbell, Phys. Rev. E {\bf 48}, 3077 (1993).

\bibitem{wright}
O. Zobay {\em et. al.},
Phys. Rev. A {\bf 59}, 643 (1998);
P. Meystre, {\em Atom Optics} (Springer, NY, 2001), Ch. 11.

\bibitem{markus}
K.M. Hilligsøe, M.K. Oberthaler, and K.P. Marzlin
Phys. Rev. A {\bf 66}, 063605 (2002).


\bibitem{kivshar}
P.J.Y. Louis {\em et. al.}, Phys. Rev. A {\bf 67}, 013602 (2003).

\bibitem{scott}
R.G. Scott {\em et. al.}, Phys. Rev. Lett. {\bf 90}, 110404 (2003).

\bibitem{bloch_book}
N.W. Ashcroft and N.D. Mermin,
{\em Solid State Physics}
(Saunders College, NY, 1976).

\bibitem{ben}
{\em Nonlinear Photonic Crystals}, R.E. Slusher and B.J. Eggleton, Eds., (Springer, 2003).


\bibitem{landau}
L.D. Landau, E.M. Lifshits,
{\em  Electrodynamics of Continuous Media} (Moscow, Nauka, 1992), Ch. 116.

\bibitem{erman1}
Experimental control of the matter-wave dispersion using this
technique has been recently demonstrated in Ref. \cite{erman}.

\bibitem{mario}
V.V. Konotop and M. Salerno, Phys. Rev. A {\bf 65}, 021602 (2002).



\bibitem{pismen}
L.M. Pismen, {\em Vortices in Nonlinear Fields} (Clarendon Press, Oxford, 1999), Ch. 4.

\bibitem{adams}
N.G. Parker {\em et al.}, Phys. Rev. Lett. {\bf 90}, 220401 (2003).

\end{references}
\end{document}